\documentstyle[epsfig]{mn}

\newif\ifAMStwofonts

\ifoldfss
  \newcommand{\rmn}[1] {{\rm #1}}

  \ifCUPmtlplainloaded \else
    \NewTextAlphabet{textbfit} {cmbxti10} {}
    \NewTextAlphabet{textbfss} {cmssbx10} {}
    \NewMathAlphabet{mathbfit} {cmbxti10} {} 
    \NewMathAlphabet{mathbfss} {cmssbx10} {} 
  \fi
  \ifAMStwofonts
    \ifCUPmtlplainloaded \else
      \NewSymbolFont{upmath} {eurm10}
      \NewSymbolFont{AMSa} {msam10}
      \NewMathSymbol{\upi}     {0}{upmath}{19}
      \NewMathSymbol{\umu}     {0}{upmath}{16}
      \NewMathSymbol{\upartial}{0}{upmath}{40}
      \NewMathSymbol{\leqslant}{3}{AMSa}{36}
      \NewMathSymbol{\geqslant}{3}{AMSa}{3E}

    \fi
  \fi
\fi 

\ifnfssone
  \newmathalphabet{\mathit}
  \addtoversion{normal}{\mathit}{cmr}{m}{it}
  \addtoversion{bold}{\mathit}{cmr}{bx}{it}
  \newcommand{\rmn}[1] {\mathrm{#1}}

  \newmathalphabet{\mathbfit} 
  \addtoversion{normal}{\mathbfit}{cmr}{bx}{it}
  \addtoversion{bold}{\mathbfit}{cmr}{bx}{it}
  \newmathalphabet{\mathbfss} 
  \addtoversion{normal}{\mathbfss}{cmss}{bx}{n}
  \addtoversion{bold}{\mathbfss}{cmss}{bx}{n}
  \ifAMStwofonts
    \ifCUPmtlplainloaded \else
      \UseAMStwoboldmath
      \makeatletter
      \new@mathgroup\upmath@group
      \define@mathgroup\mv@normal\upmath@group{eur}{m}{n}
      \define@mathgroup\mv@bold\upmath@group{eur}{b}{n}
      \edef\UPM{\hexnumber\upmath@group}
      \new@mathgroup\amsa@group
      \define@mathgroup\mv@normal\amsa@group{msa}{m}{n}
      \define@mathgroup\mv@bold\amsa@group{msa}{m}{n}
      \edef\AMSa{\hexnumber\amsa@group}
      \makeatother
      \mathchardef\upi="0\UPM19
      \mathchardef\umu="0\UPM16
      \mathchardef\upartial="0\UPM40
      \mathchardef\leqslant="3\AMSa36
      \mathchardef\geqslant="3\AMSa3E
    \fi
  \fi
\fi 

\ifnfsstwo
  \newcommand{\rmn}[1] {\mathrm{#1}}

  \DeclareMathAlphabet{\mathbfit}{OT1}{cmr}{bx}{it}
  \SetMathAlphabet\mathbfit{bold}{OT1}{cmr}{bx}{it}
  \DeclareMathAlphabet{\mathbfss}{OT1}{cmss}{bx}{n}
  \SetMathAlphabet\mathbfss{bold}{OT1}{cmss}{bx}{n}
  \ifAMStwofonts
    \ifCUPmtlplainloaded \else
      \DeclareSymbolFont{UPM}{U}{eur}{m}{n}
      \SetSymbolFont{UPM}{bold}{U}{eur}{b}{n}
      \DeclareSymbolFont{AMSa}{U}{msa}{m}{n}
      \DeclareMathSymbol{\upi}{0}{UPM}{"19}
      \DeclareMathSymbol{\umu}{0}{UPM}{"16}
      \DeclareMathSymbol{\upartial}{0}{UPM}{"40}
      \DeclareMathSymbol{\leqslant}{3}{AMSa}{"36}
      \DeclareMathSymbol{\geqslant}{3}{AMSa}{"3E}
    \fi
  \fi
\fi 

\ifCUPmtlplainloaded \else
  \ifAMStwofonts \else 
    \def\upi{\pi}
    \def\umu{\mu}
    \def\upartial{\partial}
  \fi
\fi

\title{The characteristic stellar mass as a function of redshift}
\author[C. J. Clarke and V. Bromm]
       {Cathie J. Clarke$^{1}$ and Volker Bromm$^{1,2}$ \\
      $^1$Institute of Astronomy, Madingley Road, Cambridge CB3 0HA \\
$^2$Harvard-Smithsonian Center for Astrophysics, 60 Garden Street, Cambridge, MA 02138, U.S.A.}

\pagerange{\pageref{firstpage}--\pageref{lastpage}}
\pubyear{2002}

\begin{document}

\maketitle

\label{firstpage}

\begin{abstract}
We present a model for the star formation process during the initial
collapse of dark matter haloes at redshifts $z=0-30$. We derive a simple
formula for the characteristic stellar mass scale
during this initial burst of star
formation. In our picture, this characteristic scale reflects both
the minimum temperature to which the gas can cool (determined by
the metallicity and the temperature of the cosmic microwave background)
and the pressure of overlying baryons in the collapsing halo. This
prescription reproduces both the large mass scales found in simulations
of Population~III star formation and the near solar values observed for
star formation at low redshift. 
\end{abstract}

\begin{keywords}
cosmology: theory -- early universe -- galaxies: formation -- 
stars: formation.
\end{keywords}

\section{Introduction}

  It has long been recognised that the observed form of the
initial mass function (IMF)
in the Milky Way imprints the stellar population with
a characteristic mass somewhat below $1 M_{\odot}$. The
existence of a characteristic scale, linked to neither the upper
nor lower limits of the observed IMF, owes itself to the observed
flattening of the mass function in this region (Scalo 1986, 1998;
Kroupa, Tout \& Gilmore 1990). If the IMF is parameterised
by a series of disjoint power laws with respective indices $\alpha$
(such that the number of stars in the mass range $m$ to $m+{\rmn d}m$ is
$\propto m^{-\alpha}{\rmn d}m$), then the total mass contributed by each
power law section is dominated by the upper (lower) mass limits
of the section if $\alpha$ is respectively $<$ ($>$) $2$. Since
$\alpha $ makes the transition from $ \sim
2.35$ to $\sim 1.5$ at around a solar mass, it follows that the characteristic
stellar mass is around this value. 

  At higher masses, the IMF extends in a  power law
for around two orders of magnitude in mass (Salpeter 1955). It is tempting
to speculate on the origin of the two main features of the IMF
(see Larson 1995, 1996, 1998a).The slope of the upper IMF power law 
may be set by some scale free process.
In this regard,
both coagulation and competitive accretion have been cited as possible 
mechanisms
(e.g. Murray \& Lin 1996; Bonnell et al. 2001).
The characteristic mass scale, on the other hand, is argued to be set by some
physical property of the star forming gas and, therefore, may be expected
to vary with environment and epoch (e.g. Larson 1998b).

  In nearby star forming regions, the characteristic stellar
mass is similar to the typical Jeans mass in dense molecular
cloud cores. This Jeans mass is jointly set by the
gas temperature and the pressure in the cores. Whereas the temperature
is fixed by the detailed physics of heating and cooling
in molecular gas, 
the thermal pressure
in the cores is apparently in a state of rough pressure
balance with the mean internal pressure of bulk motions
(`turbulence') within the parent molecular cloud (Larson 1996, 1998a).
This conjecture
has subsequently been
supported by hydrodynamic simulations of turbulent clouds.
These numerical experiments
show that the density field generated by driven supersonic
turbulence in an isothermal medium gives rise to a characteristic
density at which the thermal pressure roughly balances the
turbulent pressure in the cloud 
(e.g. Padoan, Nordlund \& Jones 1997; Padoan \& Nordlund 2002).
Through analysis of the cloud's
steady state density distribution, and by converting density
to corresponding isothermal Jeans mass, these authors derive a
characteristic mass and (approximately log-normal) IMF for the
cloud. 
Similar conclusions may
be derived from the recent simulations of Bate, Bonnell \&
Bromm (2002a,b, 2003). This work differs from the above both in that the
turbulence is not driven (but allowed to decay on roughly
a cloud free-fall time) and also, more crucially, in that
it follows the collapse and fragmentation of gravitationally
unstable gas down to the opacity limit and hence directly
simulates the building up of the IMF by combined fragmentation
and competitive accretion. In this case also, the mean
stellar mass appears to be set by the gas temperature and the turbulent
pressure of the initial conditions.
For observed molecular clouds with the typical internal pressure 
($nT \sim 3 \times 10^5$K cm$^{-3}$) and temperature ($\sim 10$\,K)
of molecular gas cooled by CO line emission
the characteristic stellar  mass is around a solar mass.

   In order to extend this argument so as to predict the characteristic
mass scale for star formation at other epochs, and in different
environments, it is evidently necessary to
consider the factors
determining  both the gas temperature and the mean pressure within
star forming complexes. The former issue
has been considered by numerous authors with the well known result that
higher mass stars are to be expected at high redshift, both due
to inefficient cooling of primordial gas and also to the raising of
the floor set by the temperature of the cosmic microwave background
(e.g. Schwarzschild \& Spitzer 1953; Larson 1986, 1998b). The
role of pressure variations has received less attention, partly because
of the much weaker (inverse square root) dependence of the Jeans mass on 
pressure compared with that on temperature. 

  In nearby molecular clouds,
two factors appear to be relevant in setting the pressure. First, it  is
evident that
the attainment of the low temperatures needed for star formation
requires that molecular coolants are self-shielded against
photodissociation by the ambient UV field (e.g. van Dishoeck \&
Black 1988). This imposes
a minimum column density for molecular clouds of around $ 3 \times
10^{21}$ cm$^{-2}$.
In a self-gravitating system in which the flow velocities are
comparable to their free-fall values, the pressure depends
on the column density according to $P \sim G \Sigma^2$
and thus the self-shielding requirement  
translates into a minimum viable pressure of around 
$2 \times 10^4$K cm$^{-3}$.
On the other hand, it is
also apparent that the ambient pressure of the interstellar medium (ISM)
plays a role in setting a lower limit to the mean internal pressure. 
Local giant molecular clouds (GMCs) are somewhat
self-gravitating and the internal pressure of molecular clouds thus
exceeds the ambient pressure
by around an order of magnitude. The pressure within GMCs (or,
equivalently, their column densities) does however appear to scale with
the  ambient interstellar pressure, in that the  pressures within
GMCs in the Galactic centre are significantly higher than those in local
clouds (Sanders, Scoville \& Solomon 1985). 
In the following, we hypothesise that the pressure
in star forming environments
modestly exceeds that of the ambient ISM. 

 Whereas we here focus on possible variations in the stellar IMF,
we do not address the history of the cosmic star formation rate (SFR).
This problem has been investigated by many authors, both
analytically and numerically (e.g. Norman \& Spaans 1997;
Barkana \& Loeb 2000; Springel \& Hernquist 2003). A physical understanding
of the cosmic SFR crucially depends on the nature of the feedback
exerted by star formation on its surroundings. The character of this
feedback is in turn determined by the underlying IMF. The results
from our study therefore provide an important ingredient to the
overall effort to elucidate the star formation history of the universe.

 In this paper, we estimate the characteristic stellar mass associated with
bursts of star formation accompanying the formation of galaxies. To this
end, we consider the typical gas pressure and temperature
in protogalaxies as they collapse
and virialise.
In the context of a hierarchical model of cosmic structure formation, we
evaluate how the characteristic stellar mass varies as a function of
collapse redshift and halo mass. Specifically, we
assume a $\Lambda$CDM cosmology with density parameters in matter
$\Omega_{m}=1-\Omega_{\Lambda}=0.3$, and in baryons $\Omega_{\rmn B}=0.045$,
a Hubble constant of $h=H_{0}/100$ km s$^{-1}$ Mpc$^{-1}=0.7$, and a 
scale-invariant power spectrum of density fluctuations with an amplitude
$\sigma_{8}=0.9$ on a scale of 8 $h^{-1}$ Mpc.

\section{Star Formation Model}

  In what follows, we adopt the view that the internal pressure in star
forming systems at different cosmological epochs is approximately determined
by the ambient pressure in the ISM. 
In particular, we will be
focusing on the bursts of star formation that accompany the first
collapse of gas into dark haloes during the assembly of galaxies. We use
insights 
from recent numerical simulations
into the thermodynamic behaviour of gas in dark haloes virialising
at high redshifts
(Bromm, Coppi \& Larson 1999, 2002; Abel, Bryan \& Norman 2002).

  As overdensities in the dark mater distribution turn around from
the mean cosmic expansion  and collapse, the baryons they contain
are initially heated by adiabatic compression. For a halo virialising
at redshift $z$, the dark matter density at the point of virialisation
is:

\begin{equation}
\rho_{\rm vir} \simeq 200\rho_{b}(1+z)^{3}\mbox{\ ,}
\end{equation}
where
\begin{displaymath}
\rho_{b}=2\times 10^{-29}h^{2}\Omega_{m}\mbox{\ g cm$^{-3}$}
\end{displaymath}
is the density of the background Universe.
Assuming a cosmic ratio of baryons to dark matter of
$0.15$, the baryonic density at this point is:

\begin{equation}
 n_{1} \simeq 0.3 {\rm cm}^{-3} \left(\frac{1+z_{\rmn vir}}{20}\right)^{3}
\mbox{\ .}
\end{equation}
At this stage the temperature attained through adiabatic compression, $T\sim
500{\rm \,K}$, is considerably less than the virial temperature of the halo (see
Barkana \& Loeb 2001)
\begin{equation}
T_{\rm vir} \simeq 2000 {\rm \,K\,} M_{6}^{2/3}\left(\frac{1+z_{\rm vir}}{20}
\right){\mbox \, ,}
\end{equation}
where $M_{x}=M/10^{x}M_{\odot}$.
The baryons, not being pressure supported in such a well, thus contract
and continue to heat up.  The further evolution depends on the mass 
of the halo. In the case of low mass haloes, the gas can attain a temperature
of $\sim T_{\rmn vir}$ by adiabatic compression alone. 
The corresponding density
can readily be estimated by extrapolating the adiabat from the
thermodynamic state at $z=100$ ($n \simeq 0.1 {\rm cm}^{-3}$,
$T\simeq 200 {\rm K}$)
to $T_{\rmn vir}$:
\begin{equation}
n_{2} \simeq 3 {\rm cm}^{-3} M_{6} \left(\frac{1+z_{\rmn vir}}{20}\right)^{3/2}
\end{equation}

  A different evolutionary path is followed in the case of high mass
haloes, however, since centrifugal effects become important in
halting radial collapse. We may estimate the halo mass at which
this occurs by equating the maximum density attained in quasi-spherical
collapse ($\sim n_1/\lambda^3$, where $\lambda$ is the usual spin parameter)
to $n_2$ derived above, and find a critical halo mass of
\begin{equation}
M_{c} \simeq 10^{8} M_{\odot} \left(\frac{1+z_{\rmn vir}}{20}\right)^{3/2} \lambda_{0
.1}^{-3}\mbox{\ .}
\end{equation}
Here, the spin parameter is normalised to a fiducial value of 0.1,
close to what is found in cosmological simulations (e.g. Padmanabhan 1993).
In such massive haloes, 
once radial contraction is slowed by centrifugal support, shocks develop
as infalling gas joins the incipient disc. The introduction of entropy in
shocks causes the evolution in the $[n,T]$ plane
to steepen with respect to the slope of adiabatic evolution, and we
estimate the approximate postshock parameters to be: $n \simeq n_1/\lambda^3$ and
$T \simeq T_{\rmn vir}$.

   Thus, to summarise, prior to the onset of cooling, the gas arrives
at a state with temperature $T_v=T_{\rmn vir}$
and density $n_v = {\rm min} \biggl(n_2,n_1/\lambda^3\biggr)$, where the relevant
density depends on whether the halo mass is less
than or greater than $M_c$. For $M<M_c$, rotational support is
negligible in the virial state, whereas in the opposite limit, rotation,
thermal and gravitational effects are all comparable prior to the onset of
cooling.

  The subsequent evolution depends on the efficiency of cooling at
conditions corresponding to $n_v,T_{v}$.
In general, the gas in a virialised dark matter (DM) halo will continue
to collapse and fragment if the criterion $t_{\rmn cool} < t_{\rmn ff}$
is satisfied (Rees \& Ostriker 1977).
Systems with halo masses
in excess of $\sim 10^6 M_\odot$ fulfil the Rees-Ostriker criterion
either if H$_2$ cooling is effective (implying the
absence of a sufficiently strong photodissociating UV field) or else,
in the absence of photoionisation heating,  if
the metallicity exceeds $\simeq 10^{-3.5} Z_\odot$ (Omukai 2000; Bromm et al.
2001). More massive systems, with $M\ga 10^{8}M_{\odot}[(1+z)/10]^{-1.5}$,
are able to collapse via atomic hydrogen lines (e.g. Madau, Ferrara, \& Rees 2001;
Oh \& Haiman 2002).
In each case, we assume that the gas is compressed roughly {\it isobarically}
as it cools. Below, we discuss how the results of numerical simulations lend
support 
to this idealized model. 
During the isobaric phase the pressure is
\begin{eqnarray}
P_{v} & \simeq & 10^{7}k_{\rm B}\times {\rm min} (M_{8}^{5/3}
\left(\frac{1+z_{\rmn vir}}{20}\right)^{5/2}, \nonumber \\
 & & 2 M_{8}^{2/3} \left(\frac{1+z_{\rmn vir}}{20}\right)^{4}
 \lambda_{0.1}^{-3}) {\rm \ K\, cm}^{-3}
\end{eqnarray}

A similar dependence of gas pressure on mass and redshift compared
to our high-mass case has been found by Norman \& Spaans (1997).
These authors have studied the properties of protogalactic discs,
and it is therefore not surprising that our high-mass case, where
centrifugal support is important, leads to a similar overall scaling.

 We assume that the immediate progenitor of a star is a centrally concentrated,
self-gravitating core (Motte, Andr\'{e}, \& Neri 1998) with a mass
close to the Bonnor-Ebert value (e.g. Palla 2002)
\begin{equation}
M_{\rm BE}\simeq 700 M_{\odot} \left(\frac{T}{200 {\rm \,K}}
\right)^{3/2}\left(\frac{n}{10^{4}{\rm cm}^{-3}}\right)^{-1/2} \mbox{\ .}
\end{equation}

To evaluate the minimum possible fragment mass, we need to determine
$T_{\rm min}$ and $n_{\rm max}$.
The minimum temperature is given by
\begin{equation}
T_{\rm min} = \mbox{max}(T_{\rm cool},T_{\rm CMB})\mbox{\ ,}
\end{equation}
where
\begin{equation}
T_{\rm cool} = \left\{
\begin{array}{ll}
\mbox{200~K} & \mbox{for H$_{2}$}\\
\mbox{10~K} & \mbox{for CO}\\
\end{array}
\right. \mbox{\ .}
\end{equation}
Note that although throughout the paper we refer to the case that
temperatures of $\sim 10$~K are obtained as `CO' cooling, dust may
also provide a means of achieving similarly low temperatures even in the
absence of molecular coolants (see, e.g. Whitworth, Boffin \& Francis 
1998). 
A lower floor on the gas temperature is set by the cosmic
microwave background (CMB) with
\begin{equation}
T_{\rm CMB} = \mbox{2.7~K}(1+z)\mbox{\ .}
\end{equation}
We find the maximum density by assuming that the
dissipative infall
proceeds along an isobar, resulting in
\begin{equation}
n_{\rm max} \simeq P_{v} /(k_{\rm B} T_{\rm min}) \mbox{\ .}
\end{equation}
Inserting these expressions for $T_{\rm min}$ and $n_{\rm max}$
into equ. (7), we find for the low-mass case
\begin{equation}
M_{\rm BE}\simeq  200 M_{\odot} \left(\frac{T_{\rm min}}{{\rm 200~K}}
\right)^{2} M_{8}^{-5/6}
\left(\frac{1+z_{\rm vir}}{ 20}\right)^{-5/4} \mbox{\ ,}
\end{equation}
and for the high-mass case
\begin{equation}
M_{\rm BE}\simeq 140 M_{\odot} \left(\frac{T_{\rm min}}{{\rm 200~K}}
\right)^{2} M_{8}^{-1/3}
\left(\frac{1+z_{\rm vir}}{ 20}\right)^{-2}\lambda_{0.1}^{3/2} \mbox{\ .}
\end{equation}

Depending on the details of how the gas is accreted onto the nascent
protostar in the centre of the collapsing core, the resulting stellar mass
is expected to be somewhat smaller than the Bonnor-Ebert mass.
We take this uncertainty into account
by expressing the final characteristic mass as
\begin{equation}
M_{\rmn char} \simeq \alpha M_{\rm BE} \mbox{\ ,}
\end{equation}
and choose the efficiency to be $\alpha\simeq 0.5$.
This value is close to that inferred for the formation of stars in the
present-day Universe (McKee \& Tan 2002).
This choice of efficiency reproduces the high redshift (Pop~III)
point, estimating the typical mass of such a star to be a few hundred $M_{\odot}
$,
collapsing in a halo of total mass $\sim 10^{6}M_{\odot}$ at $z\sim 20$
(Bromm et al. 1999, 2002;
Nakamura \& Umemura 2001; Abel et al. 2002).

\begin{figure}
\vspace{2pt}
\epsfig{file=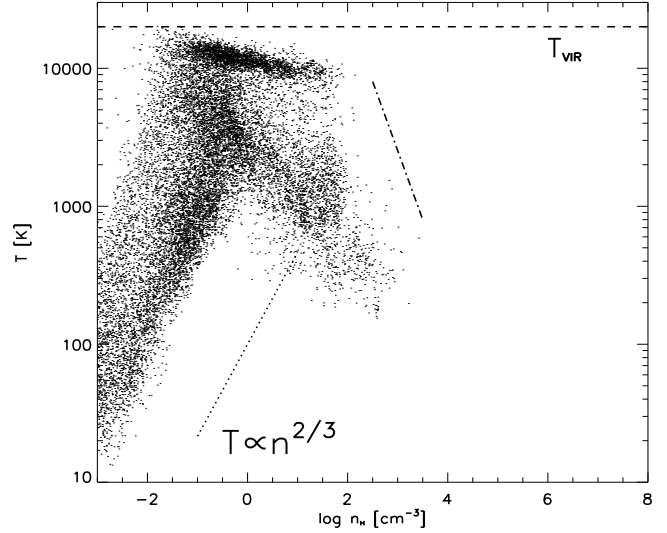,width=8.5cm,height=7.cm}
\caption{Thermodynamic behaviour of primordial (metal-free) gas in
a dwarf-sized system.
The dark halo has a total mass of $M=10^{8}M_{\odot}$ and
collapses at $z_{\rmn vir}\simeq 10$. Shown is the gas temperature
vs. hydrogen number density. At low gas densities, the temperature
rises because of adiabatic compression and due to shocks until it
reaches the virial value, $T_{\rmn vir}\simeq 2\times 10^{4}$K. At higher
densities,
H$_{2}$ line cooling drives the temperature down again to $T\sim 200$\,K.
{\it Dot-dashed line:} Evolution along an isobar.
Notice that the cooling flow at high densities asymptotically approaches
an isobaric path.
}
\end{figure}

  Clearly, this model is highly idealised, and serves only to provide
an order of magnitude estimate for typical densities and pressures
to be expected in regions of gas collapsing in proto-galactic
potentials. Some support for this schematic evolution in the $[n,T]$
phase diagram is provided by the results of numerical simulations
of the formation of dwarf galaxies at high redshifts. For example, Figure 1
depicts the phase diagram for gas collapsing  in a dark halo
of mass $10^8 M_\odot$ that virialises at $z \sim 10$. This calculation
is similar to that described in Bromm \& Clarke (2002), except that
in the case presented here 
it is assumed that molecular
hydrogen is not photodissociated, and that
the gas is metal-free. The only
effective coolant at temperatures below $10^4$K is then 
molecular hydrogen (e.g. Tegmark et al. 1997).

  Figure 1 shows that the evolution can be loosely described
as adiabatic compression followed by roughly isobaric compression from
the virialised state. In fact, since the halo mass in this
case is quite close to the transition mass in equation (5), one
can already see that shock heating is of some importance prior to
virialisation, causing the path in the $[n,T]$ plane to be
somewhat steeper than an adiabat at this stage.
Following virialisation, the gas approximately evolves along an isobar.
Note that this gas is not
initially self-gravitating but is pressure confined by the weight of
the overlying baryons.
This roughly isobaric behaviour can be understood as follows: although
the cooling timescale, $t_{\rmn cool}$,  is less than the free-fall
timescale, $t_{\rmn ff}$, it is longer than the sound-crossing
time. Thus there is sufficient time, as the gas cools, for the
passage of sound waves to maintain roughly isobaric conditions.
Such an isobaric evolution has been predicted for 
pregalactic shocks that are able to cool via H$_{2}$ (e.g. Shapiro
\& Kang 1987).
 Applying our model to the system described in Figure 1, the predicted
density at the minimum temperature
(200~K) is around $100$ cm$^{-3}$, in fair agreement with the simulation.

Our simple model is in effect a one-zone model, whereas the gas in
a given dark matter halo will have a pressure and density that is dependent
on radius. To constrain the possible influence of a radial pressure profile
on the predicted characteristic stellar mass, we have evaluated the
gas pressure in the simulation of Fig.~1. We find that the pressure is
roughly independent of radius within the central $\sim 150$~pc, and drops
by one order of magnitude out to a radius of $\sim 400$~pc. The 
characteristic stellar mass in the center of the incipient
dwarf galaxy, therefore, varies only within a factor of a few due to the
radial drop in pressure. We emphasize that our model is not able
to reliably assess possible spatial variations in the IMF, and a more
sophisticated approach would be needed to do so.

  The above argument (that the pressure in the proto-galaxy is set
by the weight of overlying baryons) can however be generalised
beyond the situation depicted in Figure 1. In this simulation,
the gas is never in a state of hydrostatic equilibrium: gas in
the inner regions that can cool responds to the weight of overlying
gas by increasing its density to match the pressure imposed from
above. We note that any bulk (`turbulent') motions induced in the
collapse that are gravitational in origin (e.g. Abel et al. 2002)
will introduce a ram pressure whose magnitude is at most
comparable with this gravitational pressure. Likewise, in any galaxy whose gas  subsequently attains a state of
hydrodrostatic equilibrium, the central pressure (or, in a disc
system, pressure in the mid-plane) adjusts to a similar value,
regardless of the microphysics of the processes providing
this pressure.
For example, in the disc of the Milky Way (where the ambient
pressure has significant contributions from cosmic rays, magnetic fields
and thermal pressure) the {\it value} of this pressure
($2.8 \times 10^4$ K cm$^{-3}$) is simply given by the requirement
that it supports the weight of the ISM. We have  argued in Section~1
that in any
galaxy this ambient pressure sets a lower limit to the internal
pressure of self-gravitating, potentially star forming,  clouds and that
this pressure affects the Jeans mass of stars forming from
cooled gas in these clouds.

\begin{figure}
\vspace{2pt}
\epsfig{file=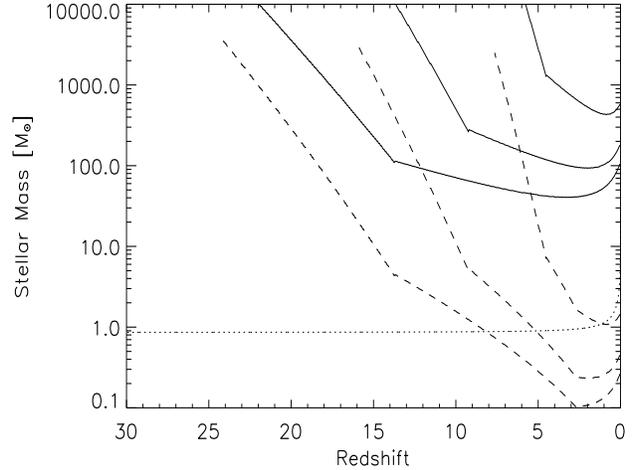,width=8.5cm,height=7.cm}
\caption{Evolution of characteristic stellar  mass in collapsing DM haloes.
Shown is the characteristic mass (in units of $M_{\odot}$) for various
overdensities vs.
redshift.
{\it Solid lines:} Cooling is due to H$_{2}$ with a minimum temperature
of 200~K.
{\it Long-dashed lines:} Cooling is due to CO with a minimum temperature
of 10~K. 
For each assumption on the cooling, the curves correspond (from
top to bottom) to 1$\sigma$, 2$\sigma$, and 3$\sigma$ perturbations, as
shown in Fig.~4.
{\it Dotted line:} Maximum mass of a star formed at a redshift $z$ that
would have survived to the present time.
}
\end{figure}

\begin{figure}
\vspace{2pt}
\epsfig{file=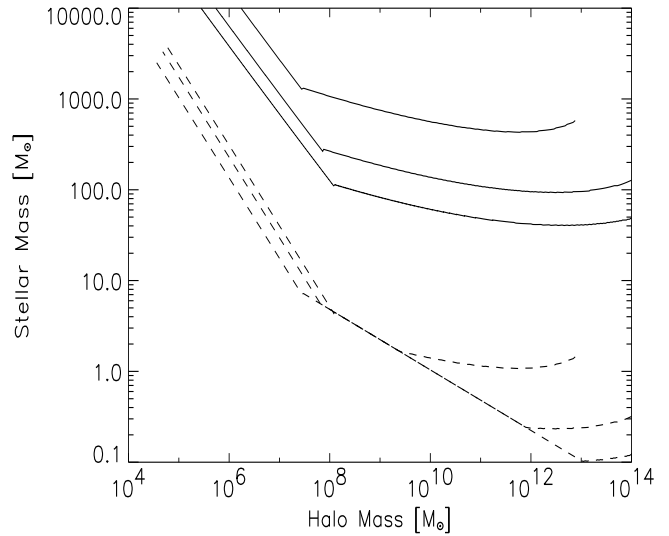,width=8.5cm,height=7.cm}
\caption{Evolution of characteristic stellar mass in collapsing DM haloes.
Shown is the characteristic mass (in units of $M_{\odot}$) for various
overdensities vs.
total (DM $+$ gas) halo mass (also in units of $M_{\odot}$).
For the different lines, we adopt the same convention as in Fig.~2.
}
\end{figure}

\begin{figure}
\vspace{2pt}
\epsfig{file=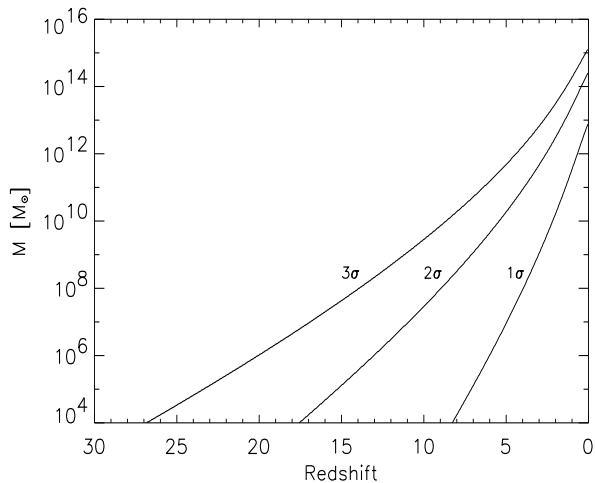,width=8.5cm,height=7.cm}
\caption{Gravitational evolution of cold dark matter haloes.
Shown is the total halo mass for various overdensities vs. redshift.
The curves are calculated for a $\Lambda$CDM spectrum with $\sigma_{8}=0.9$
and $h=0.7$.
}
\end{figure}

\section {Results}

 In Figure 2 we plot the predicted dependence of characteristic
stellar mass on the redshift of virialisation of the parent halo.
Each line represents the characteristic stellar mass for haloes
virialising at a given redshift that lie on a locus of constant 
overdensity in the random
field of primordial density perturbations. In order to deduce the
mass of the parent halo in which such stars are formed, Figure 2 should
be examined in conjunction with Figure 4, which depicts the relationship
between halo mass and virialisation redshift along lines of constant
overdensity for a $\Lambda$CDM model. 

  The solid lines represent 
the case that gas cannot cool below  $200$~K, thus approximating the
case of cooling by molecular hydrogen. The solid lines, therefore, are
applicable to haloes which are both low metallicity and are not
subject to a strong background of photons in the Lyman-Werner
range which can dissociate molecular hydrogen. In each case, the dependence
of characteristic stellar mass on redshift flattens for haloes
more massive than about $10^{8} M_{\odot}$ (see equ. (5)). The reason
for this change is that for lower mass haloes gas is compressed
nearly adiabatically to the virial state; as the redshift of virialisation
declines, the halo mass increases steeply (Figure 4) and the strongly
increasing pressure manifests itself as a marked decline in the
characteristic stellar mass with decreasing redshift. For haloes
more massive than about $10^8 M_{\odot}$, however, the role of angular
momentum in halting the collapse weakens the mass dependence of
the virial pressure (equ. (6)), and the characteristic stellar
mass declines more gently with decreasing redshift. The ordering
by mass of the curves for various overdensities (i.e. higher charactersitic
masses for lower $\sigma$ fluctuations) is simply a result of the lower
pressure attained, at a given redshift, in the smaller mass haloes
found in lower $\sigma$ fluctuations. 

  The dashed lines correspond to the case that the gas can cool to
the maximum of the ambient CMB temperature and a temperature of $10$K,
this latter representing a fiducial value for cooling by CO and dust
in Population I regions not subject to strong radiative heating
by massive stars. These lines are all terminated at the high redshift
end at the point where the predicted characteristic stellar mass is
comparable to the total baryonic mass of the halo concerned, and
are evidently only applicable well to the right of these points. 
Unlike the case of molecular hydrogen cooling described above, the
temperature along these curves is set by the CMB for all $z > 2.7$.
The decline in temperature towards the present epoch accounts for
the steeper decline in characteristic stellar mass with decreasing
redshift than in the $H_2$ cooling case, although again there is
a mild flattening of the curves for haloes more massive than
$10^8 M_\odot$. At low redshift ($z < 2.7$), the temperature attains
a plateau value of $10$K. At these low redshifts, the virial pressure
{\it declines} with decreasing redshift, because the explicit
redshift dependence in equation (6) is more significant than the
increase of halo mass with decreasing redshift. Consequently, the
characteristic stellar mass rises modestly for $z < 2.7$. 

  Figure 3 depicts the same information as a function of halo mass,
again for curves of constant overdensity and for the two different
assumed coolants. This plot shows explicitly the weaker dependence
of characteristic stellar mass on halo mass for haloes more massive
than $\sim 10^8 M_\odot$. In the case of 
CO cooling in haloes more massive than $10^8 M_\odot$, but at
$z > 2.7$,  the
characteristic stellar mass is independent of $z$, because the
$z$ dependence of the pressure (equ. (6)) and that
of the temperature of the CMB cancel each other out in
their contribution to the resulting stellar mass. Consequently,
curves of different overdensity follow the same dependence
of characteristic stellar mass on halo mass in this regime. For
each line of constant overdensity, the characteristic stellar
mass departs from this relation at a redshift of
$\sim 2.7$, where the plateau in the temperature attained causes
a modest upturn in the characteristic stellar mass.

\section {Implications}
\subsection {Summary of predicted behaviour}

  We have proposed a model that relates the masses of dark haloes
and their virialisation redshift to the characteristic mass scale
of stars, formed during the assembly of protogalaxies. 
We stress that the characteristic mass that we derive here
is neither an upper- nor lower-mass cutoff, but is instead
close to the mean mass. For an IMF of form similar to that seen
in well observed local regions, this mass scale marks the
transition between a power law index less than $2$ to a near
Salpeter ($\alpha=2.35$) value.

  This characteristic mass scale is identified with the Jeans mass,
fixed jointly by the gas temperature (set by local cooling
physics together with any floor set by the temperature of the
cosmic microwave background)  and pressure (set by the weight
of overlying baryons as the gas collapses in the parent dark halo).
This simple prescription implies a near solar characteristic
mass scale for star formation at recent epochs (in line with
observations), as well as reproducing the much higher mass
scale ($\la 10^3 M_\odot$) found in simulations of Population
III haloes where the cooling is provided mainly by molecular
hydrogen.

  At redshifts greater than around 3, the characteristic mass
scale increases with $z$, due partly to the increase in temperature
of the cosmic microwave background with $z$ and partly because, along
a line of constant overdensity $n\sigma$ (see Figure 4), the halo
mass decreases  with increasing $z$ 
and hence the pressure in the halo declines. Although the Jeans
mass varies only as $P^{-1/2}$, the steep decline in halo mass
with increasing redshift of virialisation 
ensures a significant role for pressure variations in determining
the characteristic stellar mass. The importance of pressure
variations is evident in Figure 2: each of the dashed
curves (and each of the solid curves) has an identical assumed
dependence of temperature on redshift, so that the variations
between these curves, at a given $z$, are entirely attributable
to the different masses (and hence different pressures) of the
dark haloes involved. 

\subsection {Observational consequences}

  In order to establish potentially observable diagnostics
of such a cosmic star formation model, it is necessary not
only to specify how the characteristic stellar mass varies
with cosmic epoch and conditions, but also to address the 
variation of the functional form of the IMF. 
If we assume that the form of this
IMF is invariant in different environments, but that the
value of $M_{\rmn char}$ may change, then evidently it is possible
to probe the value of $M_{\rmn char}$ only through observations
that are sensitive to masses both below and above $M_{\rmn char}$ (e.g.
Wyse et al. 2002).
In the following, we assess whether the model can be tested through
a variety of observational diagnostics.

\subsubsection {Formation of very massive black holes}

  An intriguing  application of the curves shown in Figure 2 is to the formation
of very massive black holes (VMBHs) in Population~III
haloes (Madau \& Rees 2001). Recent theoretical studies of the
thermodynamics in DM haloes prior to reionisation
have argued for a critical
metallicity threshold of around $10^{-3.5} Z_\odot$, such that cooling
to the temperature of the CMB becomes possible only in
systems that exceed this threshold (Omukai 2000; Bromm et al. 2001).
In this picture, the
attainment of the critical metallicity would cause the characteristic
stellar mass to undergo a downward transition from a given $n\sigma$
H$_{2}$-cooling curve in Figure~2 to the corresponding $n\sigma$ CO-cooling curve
at the same redshift (see also 
Mackey, Bromm \& Hernquist 2003). 
Evidently, the magnitude of the jump in characteristic mass depends on
the redshift at which this transition occurs.
As stressed by Schneider et al. (2002), the key issue here is the
fraction of the stellar population that is in the range $140-260
M_\odot$, since pair-instability supernovae from such stars provide
a very efficient way of returning metals to the ISM. The simulations
of star formation out of extremely low-$Z$ gas to date have focused on low-mass
haloes virialising at high redshift, where characteristic stellar masses
are very large. The solid lines in  Figure~2 illustrate
that at lower redshifts, the characteristic stellar mass for extremely
low-$Z$ gas should
fall, due simply to the larger pressures in the more massive
haloes virialising at this epoch. For example,    
at redshifts in the range $10-15$, the characteristic
stellar mass for primordial gas in  $2-3 \sigma$ peaks falls to a few hundred
solar masses, suggesting that the fraction of stars undergoing 
pair-production supernovae should rise at this point. We therefore
conclude that it is very unlikely that an IMF biased towards very
high masses could persist at redshifts less than $10-15$. According
to Schneider et al. (2002), this would be sufficient to account for
the density of supermassive black holes in nearby galaxies
(assuming that the very massive stars formed in this high mass
mode produced black holes that eventually merged into supermassive
black holes in galactic nuclei), but would not produce enough
remnant black holes to account for the bulk of baryonic dark matter.

\subsubsection {Element ratios}

 In metal-poor stars formed at early epochs (which have
therefore not been subject to enrichment by Type I supernovae)
the ratio of $\alpha$ elements to iron is a sensitive probe 
of the IMF in the mass range of stars contributing
Type II supernova progenitors (Tsujimoto et al. 1997). The value
of this `Type II plateau' for the [$\alpha$/Fe]-ratio
in fact constrains the IMF to have a slope very close to the
Salpeter value for stars more massive than around $8 M_\odot$. 
Inspection of Figure 2 shows that in our model, the
characteristic stellar mass is predicted to be less than
$8 M_\odot$ for $z < 15$ provided the gas can cool down to the
temperature of the CMB. The variation we predict in the
characteristic stellar mass at lower redshifts would not
leave any imprint on the [$\alpha$/Fe] element ratio from Type~II
supernovae, provided
the high mass tail of the IMF remains of Salpeter form. 

\subsubsection {Stellar populations in Lyman Break Galaxies}

 Population synthesis studies in Lyman Break Galaxies (LBGs) constrain
the IMF to be of a Salpeter form down to stars of early
B spectral type (Pettini et al. 2002).  Figure 2 demonstrates
that the characteristic stellar mass we predict at the redshift
of LBGs ($z\sim 3$) corresponds to stars of much later spectral type. Again,
therefore, the variations we predict for the characteristic 
stellar mass would have no effect on observable stellar populations
in LBGs.

\subsubsection {Frequency of very metal poor Galactic halo stars}

  The lowest metallicity stars are expected to have ages comparable with
the age of the Universe and thus are all low mass ($\la 0.8 M_\odot$). The
numbers of such stars, when combined with a chemical evolution model
and a model for the assembly of the Milky Way, can be used to constrain
the mass fraction of stars produced in this mass range at high redshift.
Such an analysis by Hernandez and Ferrara (2001) concluded that the
relative paucity of metal poor stars in the Milky Way argued for a small
mass-fraction in low mass objects. They interpreted this in terms of
a rather high $M_{\rmn char} (\sim 10 M_\odot)$ for stars originating in low
mass haloes ($10^8-10^9 M_\odot$) at a redshift of $z=5-10$. Such an
inference is in good agreement with our model (see Figures 2 and 3).  

Our model indicates that the conditions in the Universe at $z\ga 5$
have already enabled the
formation of low-mass stars, with typical masses of $\sim 1 M_{\odot}$, provided
that the gas has been sufficiently enriched with heavy elements.
Two relevant observations could thus be accomodated. First, the inferred
ages of the oldest globular clusters in the Galaxy with metallicities
of $\sim 10^{-2}Z_{\odot}$ imply a formation time close to the
epoch of reionisation (e.g. Ashman \& Zepf 2001; Cen 2001; Bromm \& Clarke 2002).
The second observational constraint is provided by the recent discovery
of an extremely metal-poor star in the Galactic halo (Christlieb et al. 2002).
The star HE0107--5240, with an iron abundance 
of [Fe/H] $= -5.3$, must have formed very early in the chemical
enrichment history of the Universe, probably out of gas that had experienced
only one previous episode of star formation. How could this low mass star have 
formed? Although this star is very iron-poor, it is dramatically overabundant
in carbon and, to a somewhat lesser extent, in oxygen and nitrogen (e.g. Schneider et al. 2003).
Taking into account the different elemental yields, it seems possible that the parent cloud
out of which HE0107-5240 formed had already a metallicity in excess of the critical
value required for low-mass star formation
(see Mackey et al. 2003 for an alternative formation mechanism).

\subsubsection {Characteristic mass scale for stars in Globular Clusters}

  The inferred age of the oldest globular clusters in the Galaxy,
when combined with current cosmological models, implies that these
systems were formed at redshifts $ z \ga 3$ (Gnedin, Lahav \& Rees 2001), and
several recent models tie their formation to an era close to
the epoch  of reionisation (Cen 2001; Bromm \& Clarke 2002). Globular
clusters represent the only systems from this era  whose IMF is
well constrained observationally. The observed mass function slope
in the range $0.1-0.8 M_\odot$ is considerably flatter than
the Salpeter slope, and thus (retaining the assumption that the
IMF is always of Salpeter form above the characteristic stellar
mass), this rules out a characteristic mass that is much less than
$0.8 M_\odot$. On the other hand, the survival of globular clusters
as bound entities rules out a very top heavy mass function ($M_{\rmn char}
\gg 1 M_{\odot}$) due to the excessive mass loss predicted from
stellar winds in this case (e.g. Kudritzki 2002).
Figure 2 demonstrates that our model predicts a characteristic 
stellar mass around $z  \sim 6$ which is roughly solar for haloes
collapsing from $2 \sigma$ peaks. Apparently, therefore, our model
predicts suitably low characteristic masses at the formation
epoch of globular clusters.

\section{ Summary and Conclusions}

This paper has set out a new framework for predicting
how the characteristic stellar mass should depend on halo mass, virialisation
redshift and chemistry. At face value, it appears broadly consistent with
available observational constraints and the results of numerical simulations
at high redshift: in particular it predicts a characteristic stellar
mass that is close to solar values at low redshifts, whilst
reproducing the very large stellar masses predicted by simulations
of Population III star formation. Our model differs from
previous discussions of the rise in characteristic stellar mass at high
redshifts in that it takes explicit account of the differing
pressures expected in haloes of different masses and virialisation
redshift. One consequence of this pressure dependence is the prediction
that an IMF biased towards very massive
values should not be able to persist at $z \la 10-15$. This is because 
the higher pressures expected in haloes virialising at lower
redshift should bring the characteristic stellar mass down to values where
copious metal production is expected via pair-instability supernovae,
and thereafter the gas should be able to cool down to the temperature
of the CMB. 

 In order to make further progress on testing this scenario against
observations, it is obviously necessary to specify not only
the variation of the characteristic stellar mass, but any variation
of the functional form of the IMF. The power law
form of the local IMF at high masses tempts us to speculate that
the upper IMF is controlled by scale free processes, and that
a roughly Salpeter tail is always to be expected at masses 
above the characteristic 
stellar mass. If this is the case, then our model is not tested by
any observational diagnostics that relate exclusively to stellar
progenitors above the characteristic stellar mass (for example
the element ratios produced by Type II supernovae would be insensitive
to the value of $M_{\rmn char}$ if this were less than $\sim 8 M_\odot$). 
Instead, the model is best tested by any diagnostics that simultaneously
probe the mass function above and below our predicted values of
$M_{\rmn char}$.  
We conclude that  the detailed observational consequences of this
framework will only emerge once
it is fully incorporated
into a hierarchical merging scenario.

\section*{Acknowledgments}

We are grateful to Martin Haehnelt and Max Pettini for helpful discussions,
as well as to Andrea Ferrara for detailed comments which improved
the presentation of this paper.
This work has been supported by the ``European Community's Research Training
Network under contract HPRN-CT-2000-0155, Young Stellar Clusters.''
VB acknowledges support by NSF grant AST-0071019. CJC gratefully
acknowledges support from the Leverhulme Trust in the
form of a Philip Leverhulme Prize.

\end{document}